\documentstyle[twocolumn,aps,pre,graphicx,amssymb,amsbsy,amsfonts]{revtex}

\begin{document}

\draft

\title{Critical behavior of propagation on small-world networks}

\author{Dami\'an H. Zanette}

\address{Consejo Nacional de
Investigaciones Cient\'{\i}ficas y T\'ecnicas\\ Centro At\'omico
Bariloche and Instituto Balseiro, 8400 Bariloche, R\'{\i}o Negro,
Argentina}

\date{\today}
\maketitle

\begin{abstract}
We  report  numerical  evidence that an epidemic-like model, which
can  be  interpreted  as  the  propagation  of  a  rumor, exhibits
critical  behavior  at  a  finite  randomness  of  the  underlying
small-world  network. The transition occurs between a regime where
the  rumor  ``dies''  in a small neighborhood of its origin, and a
regime  where  it  spreads  over  a  finite  fraction of the whole
population.  Critical  exponents are evaluated, and the dependence
of  the  critical  randomness  with  the  network  connectivity is
studied.  The  behavior  of  this  system  as  a  function  of the
small-network  randomness  bears  noticeable  similarities with an
epidemiological  model  reported recently [M. Kuperman and G.
Abramson,  Phys.  Rev.  Lett.  {\bf 86}, 2909 (2001)], in spite of
substantial differences in the respective dynamical rules.
\end{abstract}

\pacs{PACS numbers: 87.23.Ge, 89.75.Hc, 05.10.-a}

\vspace{10 pt}

Small-world  networks  have  been  introduced  as an interpolation
between  ordered  and  random  graphs,  to  capture  two  specific
features  of  real  neural,  social,  and  technological  networks
\cite{1,2}.  On  one hand, they have a relatively large clustering
coefficient,  i.e.  the  probability that two neighbors of a given
vertex  are  in  turn  mutual  neighbors  is  high,  as in ordered
graphs.  On the other, the typical separation between two vertices
is  much  smaller  than the total number of vertices, as in random
graphs.   Whereas   the   geometrical  properties  of  small-world
networks  have  been  studied  in detail \cite{1,2,3,4,5,6,7,8,9},
less attention has been paid to the dynamical properties resulting
from   their   partially  random  structure  \cite{2,9,10,11}.  As
explained  in  more  detail  below,  the  structure of small-world
networks  is  parametrized  by the randomness $p$, where $p=0$ and
$p=1$   correspond   to   fully   ordered   and   random  neworks,
respectively.  It  has  been  shown  that, in asymptotically large
small-world  networks,  statistical  geometrical properties---such
as  the mean distance between vertices---exhibit qualitatively the
same  behavior  for  any  $p>0$.  Here,  we study an epidemic-like
propagation  process  on  a small-world network which, in contrast
with geometrical  properties, shows a transition between two
qualitatively  different  dynamical  regimes  at a finite value of
$p$.  Since  this model addresses a social process and small-world
networks  are  a  presumably  realistic  representation  of social
networks,  the  analysis  could  be relevant to the description of
threshold phenomena in real societies.

Our  model  \cite{12,13}  consists  of  an  $N$-element population
where,  at  each  time  step,  each  element  adopts  one of three
possible  states.  By  analogy  with  epidemiological  SIR  models
\cite{13a},  these  states  are  called  susceptible (S), infected
(I),  and  refractory  (R). The evolution proceeds as follows. At
each  time  step  a  randomly chosen infected element $i$ contacts
another  element  $j$.  Then,  (i)  if  $j$  is in the susceptible
state,  it  becomes  infected;  (ii)  if,  on the contrary, $j$ is
infected  or  refractory,  $i$ becomes refractory. These rules are
better  interpreted  in  the  frame  of a rumor spreading process,
where  S-elements  have  not  heard the rumor yet, I-elements have
heard  the  rumor  and  are willing to transmit it, and R-elements
have  lost  their  interest  in  the rumor and do not transmit it.
Initially,  only  one  element is infected and the remaining $N-1$
elements are susceptible. During the first stage of the evolution,
the  number  of  I-elements  increases.  Since this also implies a
growth  of  the  R-population,  the contacts of I-elements between
themselves  and  with R-elements become more frequent.
After a while, in
consequence,  the  I-population  begins to decline. Eventually, it
vanishes   and   the   evolution   stops.   At   the   end,  $N_R$
elements---now  in  the  refractory  state---have been infected at
some   stage  during  the  evolution.  Numerical  simulations  and
analytical  results  show  that,  generally,  $N_R<N$. Therefore,
there  is a fraction of the population that never hears the rumor.
In  the  original version of this model, each I-element is allowed
to  contact  at random any other element of the population. It has
been proven that, in such situation, the ratio $N_R/N$ approaches
a  well-defined  limit,  $N_R/N  = 0.796\dots$, for asymptotically
large values of $N$ \cite{13b,14}.

Here, in contrast, we assume that the elements are situated at the
vertices   of   a   small-world   network,  and  contacts  can  be
established  between linked elements only. The small-world network
is   constructed  from  a  one-dimensional  ordered  network  with
periodic  boundary conditions---a ring---where each node is linked
to  its  $2K$ nearest neighbors, i.e. to the nearest $K$ neighbors
clockwise  and  counterclockwise  \cite{1,2,11}. Then, each of the
$K$  clockwise  connections  of  each  node  $i$  is  rewired with
probability  $p$  to  a randomly chosen node $j$, not belonging to
the  neighborhood  of  $i$.  A  short-cut  between  two  otherwise
distant  regions  is  thus  created. Double and multiple links are
forbidden,  and realizations where the small-world network becomes
disconnected   are  discarded.  As advanced above, the  parameter
$p$ measures the randomness  of  the  resulting  small-world
network. Note that, independently of the value of $p$, the
average number of links per site is always  $2K$. We have
performed series of $10^3$  to $10^5$ numerical realizations of
the model for several values of $p$, $N$, and  $K$.  At  each
realization, the small-world network was generated  anew  and the
evolution was recorded until the exhaustion of the I-population.

\vspace{10 pt}

\begin{figure}
\resizebox{\columnwidth}{!}{\includegraphics[angle=0]{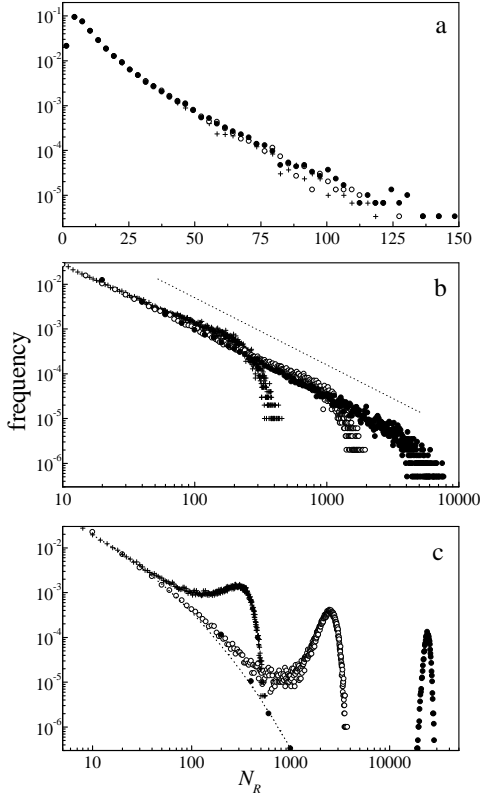}}
\caption{Frequency distribution of the number $N_R$ of R-elements
at the end of the evolution, for $K=2$ and (a) $p=0.05$, (b)
$p=0.19$, and (c) $p=0.3$. Different symbols correspond to
$N=10^3$ (crosses), $N=10^4$ (empty dots), and $N=10^5$ (full
dots). In (b), the dotted straight line corresponds to a power
law $\sim N_R^{-1.5}$. The dotted curve in (c) is a schematic
representation of the $N$-independent profile observed for small
$N_R$. Note carefully the different scales of the three plots.}
\label{f1}
\end{figure}

In  the  first  place,  we  have  studied  the distribution of the
number  $N_R$ of R-elements at the end of the evolution for $K=2$.
Figure  \ref{f1} shows the frequency $f(N_R)$ obtained from series
of  $10^5$  realizations  for selected values of $p$ and $N$. For
small  randomness  ($p=0.05$,  Fig. \ref{f1}a) the distribution is
approximately  exponential  and  does  not  depend on $N$. In this
regime,  the  rumor  ``dies''  after  a  few time steps in a small
neighborhood  of  its  origin,  due to the lost of interest of the
highly  interconnected  elements  close  to the initially infected
site.   Therefore,  the  size  $N$  of  the  whole  population  is
irrelevant  to  the  value of $N_R$. The situation is considerably
different   for   relatively   large   randomness  ($p=0.3$,  Fig.
\ref{f1}c).  Here  the  distribution  $f(N_R)$  is bimodal, with a
maximum  close  to  $N_R  =0$  (not  shown  in  the figure) and an
additional  bump for larger $N_R$. Near $N_R =0$, the frequency is
independent   of   $N$,  as  in  the  case  of  small  randomness.
Contributions   to   this  zone  of  the  distribution  come  from
realizations  where  propagation  ceases  before  a  short-cut  is
reached.   In  contrast,  the  additional  structure  is  strongly
dependent  on  $N$.  The  position  of its maximum, in fact, grows
linearly,   as   $N_{R}^{\max}\approx   0.25N$.   In   a   typical
realization  contributing  to  this zone of the distribution, many
contacts  occur  through  short-cuts  and  a finite portion of the
population  becomes infected. The intermediate regime, just before
the  large-$N_R$  structure  begins to build up, is illustrated in
Fig. \ref{f1}b  for  $p=0.19$.  The frequency follows here a power
law,  $f(N_R)  \sim N_R^{-\alpha}$ with $\alpha \approx 1.5$, over
a  substantial  interval.  This  interval is limited by above by a
smooth  cut-off,  which  shifts  to  larger  values  of  $N_R$  as
$N^\beta$, with $\beta \sim 0.5$.

\begin{figure}
\resizebox{\columnwidth}{!}{\includegraphics[angle=0]{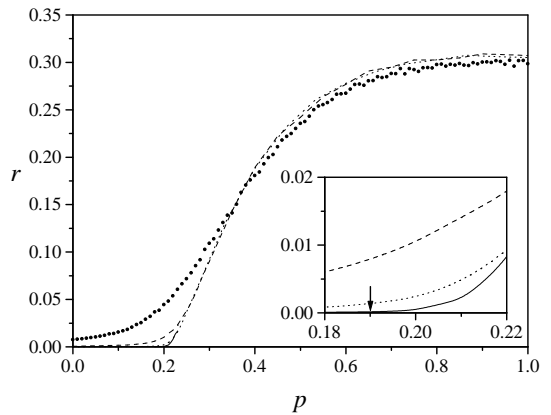}}
\caption{Average fraction $r=\langle N_R/N \rangle$
of refractory elements at the end of the evolution as a function
of the randomness $p$ on a small-world network with $K=2$, and
for $N=10^3$ (dots), $N=10^4$ (dashed line), $N=10^5$ (dotted
line), and $N=10^6$ (full line). The insert shows a close-up in
the transition zone. The arrow indicates the critical value of
$p$, determined as explained in the text.}
\label{f2}
\end{figure}

The  appearance of a well-defined power-law distribution is a clue
to  the  critical-phenomenon  nature  of  the  transition from the
regime  where  the  rumor remains localized to the regime where it
spreads  over a finite fraction of the population. To characterize
the  transition,  we  choose  as  an  order  parameter the average
fraction $N_R/N$  of R-elements at the end of the process,
\begin{equation} \label{r}
r= \langle N_R/N \rangle = N^{-1} \sum_{N_R=0}^N N_R f(N_R),
\end{equation}
which  can  be  straightforwardly  calculated  from  the numerical
results.  In  the  small-randomness  regime, where $f(N_R)$ is
independent of $N$, we expect $r\sim N^{-1}$, so that $r\to 0$ as
$N\to \infty$. For large $p$, in contrast, the presence of the
large-$N_R$ maximum in $f(N_R)$ should provide  a finite
contribution to $r$  even for asymptotically large populations.
These features are verified in numerical realizations. Figure
\ref{f2} shows $r$ as  a function of $p$, calculated in series
of  $10^3$ to $10^4$ realizations for $N$ ranging from $10^3$ to
$10^6$ and $K=2$. A well-defined transition at a finite randomness
$p_c$ is apparent for large $N$. Assuming critical behavior  at
the transition, $p_c$ can be  determined as the randomness  for
which  the extension of the power-law interval in $f(N_R)$ is
maximal. This criterion yields $p_c=0.19\pm 0.01$.

The  insert  of  Fig.  \ref{f2}  shows a close-up of the main plot
near  $p_c$.  The  smoothness of the curves, observed even for the
largest values of $N$, suggest that the critical exponent $\gamma$
associated with $r \sim |p-p_c|^\gamma$ just above the transition,
should  be  larger than unity. Linear fitting of $r$ as a function
of  $p-p_c$  in a log-log representation, for $N=10^5$ and $10^6$,
gives $\gamma \approx 2$.

\begin{figure}
\resizebox{\columnwidth}{!}{\includegraphics[angle=0]{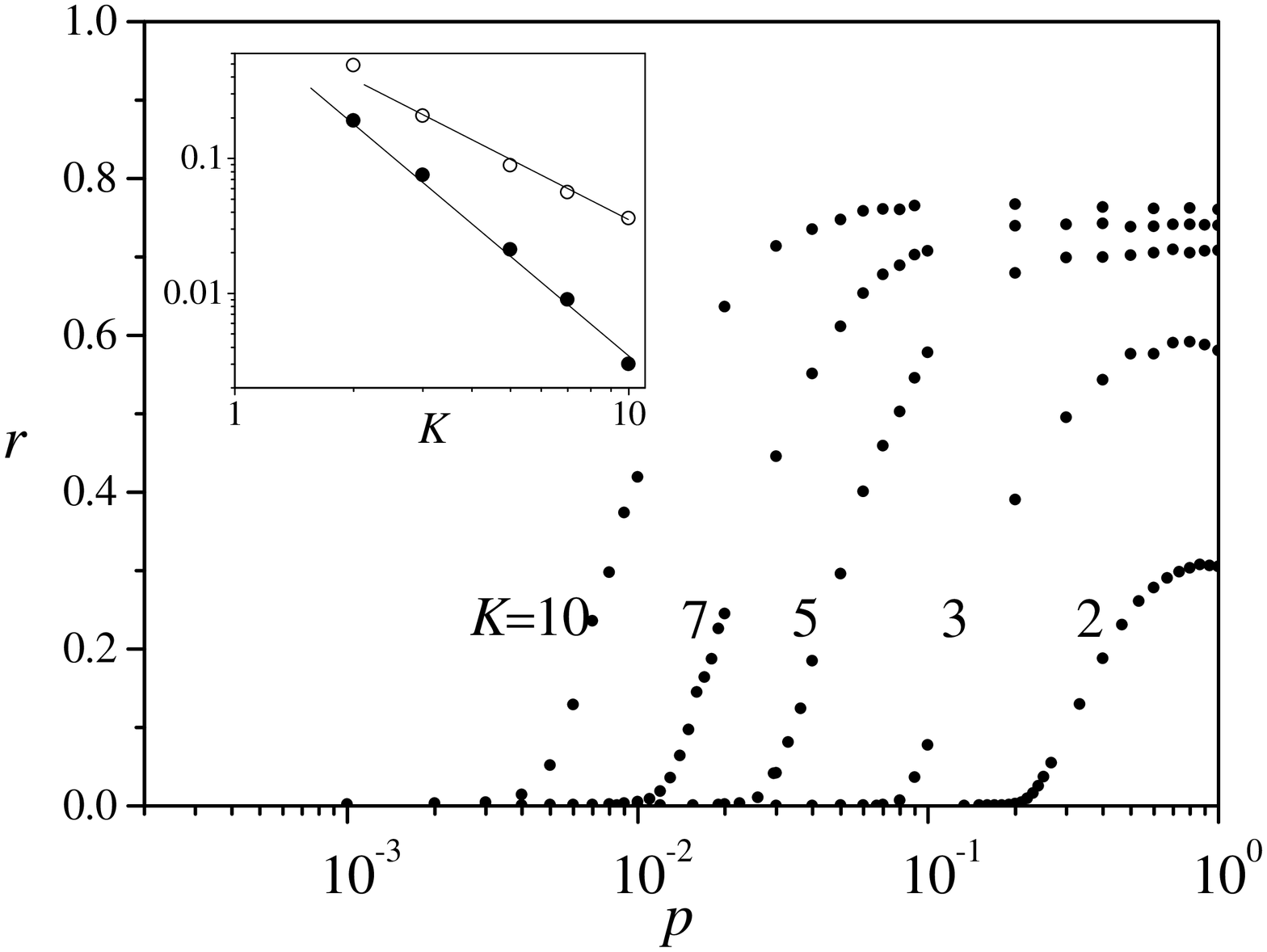}}
\caption{Average fraction $r$ of refractory elements at the end of
the evolution as a function of the randomness $p$ for several
values of $K$, and $N=10^5$.  The insert shows the critical
randomness $p_c$ (full dots) and the  difference $r^*-r_1$ (empty
dots; see text for definitions of $r^*$  and $r_1$) as a function
of $K$. Straight lines in the insert correspond to power laws
$\sim K^{-1.5}$ and $K^{-2.5}$.}
\label{f3}
\end{figure}

We have examined this model for other values of $K$, up to $K=10$,
and  found  the  same kind of transition in all cases. The average
fraction  $r$ as a function of $p$ is shown in Figure \ref{f3} for
$N=10^5$  and  several  values  of  $K$,  calculated  over  $10^3$
realizations.  It  is  seen  that  the  critical  randomness $p_c$
decreases  as  $K$  grows.  This  is  due  to the increment in the
number of long-range contacts per element, as a consequence of the
higher  connectivity of each site. On the other hand, the value of
$r$  at  $p  =  1$, $r_1$, grows with $K$ and approaches the level
expected  for  the  original  version  of  the model, $r^* = 0.796
\dots$ \cite{13b,14}.  The  insert  in  Fig. \ref{f3} displays the
dependence of
$p_c$  and  of the difference $r^*-r_1$ with $K$. Though this plot
covers  less  that  one  order  of  magnitude in the $K$-axis, the
results  suggest  power-law  decays for both quantities, $p_c \sim
K^{-\rho}$  with  $\rho  \sim 2.5$ and $r^*-r_1  \sim K^{-\sigma}$
with $\sigma \sim 1.5$.

In  summary,  our  numerical analysis shows that the regimes where
the  rumor  is  bounded  to a finite neighborhood of the initially
infected  site  and  where  it  affects  a finite fraction of the
population  are  separated by a well-defined transition at
a  finite  randomness $p_c$ of the underlying small-world network.
The results strongly suggest the presence of a critical phenomenon
at  $p_c$.  Quite  interestingly, the transition is also found if,
instead  of  building  a  frozen  small-world  network as above, a
dynamic  small-world  is considered \cite{10}. In this case, each
contact  of  an  I-element  is  established  with  one of its $2K$
nearest  neighbors  with  probability  $1-p$,  and with a randomly
chosen  site  with  probability  $p$. The  transition observed for
these  dynamic  small-worlds  is  of the  same type as on static
small-world  networks,  but  presents quantitative differences.
For $K=2$,  for  instance,  the critical  randomness  shifts  to
$p_c \approx 0.07$. Details will be discussed elsewhere \cite{fc}.

The  geometrical  properties  of  small-world  networks  have been
shown  to  exhibit a cross-over from ordered to random behavior at
$p  \sim  N^{-1}$, which implies a vanishing critical value of $p$
for  asymptotically  large  systems  \cite{5,6,7}. In contrast, we
have  here  presented  evidence that a dynamical process occurring
on  such  structures  exhibits critical behavior at a finite value
of  the  randomness.  This difference suggests that an explanation
for  the  origin  of  the  transition in geometrical terms only is
unsuitable,  and  specific dynamical properties must be taken into
account.  Very  recently, evidence of critical behavior at finite
small-world randomness  has  been reported for a purely
epidemiological model, whose  evolution  rules  are
substantially different from those    considered    here
\cite{11}.   In   particular,   the epidemiological  model
allows  for  the transformation R $\to$ S, giving  rise to a
closed disease cycle (SIRS) through recovery. Moreover, the
transformation  I $\to$ R $\to$ S is fully deterministic. The
critical  phenomenon  found in that case is a transition to global
synchronization   of   local   disease   cycles,  whereas  in  our
propagation  process  we  have  a  kind  of percolation phenomenon
\cite{8,9}.  In  spite  of these basic differences, the dependence
of  the  respective  order  parameters  on  the  randomness $p$ is
strikingly   similar,   even   if  compared  quantitatively.  This
similarity  calls  for  further investigation in order to identify
and   characterize   the  whole  class  of  small-world  dynamical
processes  with  critical  behavior  at  finite randomness, and to
give an analytical description of such behavior.

Enlightening discussions with M. Kuperman and G. Abramson are
gratefully acknowledged.


\begin{thebibliography}{20}
\bibitem{13b}
More precisely, the limit $r^*= \lim_{N\to \infty} N_R/N$ for
random contacts is the solution to the transcendental equation
$r^*=1-\exp(-2r^*)$.
\end{thebibliography}
\end{document}